\documentclass{article}

\usepackage{arxiv}

\usepackage[utf8]{inputenc} 
\usepackage[T1]{fontenc}    
\usepackage{hyperref}       
\usepackage{url}            
\usepackage{booktabs}       
\usepackage{amsmath,amssymb,amsfonts}
\usepackage{nicefrac}       
\usepackage{microtype}      
\usepackage{cleveref}       
\usepackage{graphicx}
\usepackage{natbib}
\usepackage{doi}

\usepackage{algorithmic}
\usepackage{graphicx}
\usepackage{pifont}
\usepackage{textcomp}
\usepackage{multirow}
\usepackage[table]{xcolor}
\usepackage{xspace}
\usepackage{diagbox}
\usepackage{enumitem}
\usepackage[most]{tcolorbox}
\usepackage{tikz}

\usepackage{tablefootnote}
\usepackage[ruled,linesnumbered,noend]{algorithm2e}  
\SetKwInput{KwIn}{Input}    
\SetKwInput{KwOut}{Output}
\SetKwComment{Comment}{// }{} 

\definecolor{gree}{HTML}{1e8449}
\definecolor{orang}{HTML}{b35809}

\newcommand{\titan}{\textit{TITAN}\xspace}

\usepackage{subcaption}
\usepackage{url}
\usepackage{float}

\newcommand{\paperlogo}{{\includegraphics[height=3.5em]{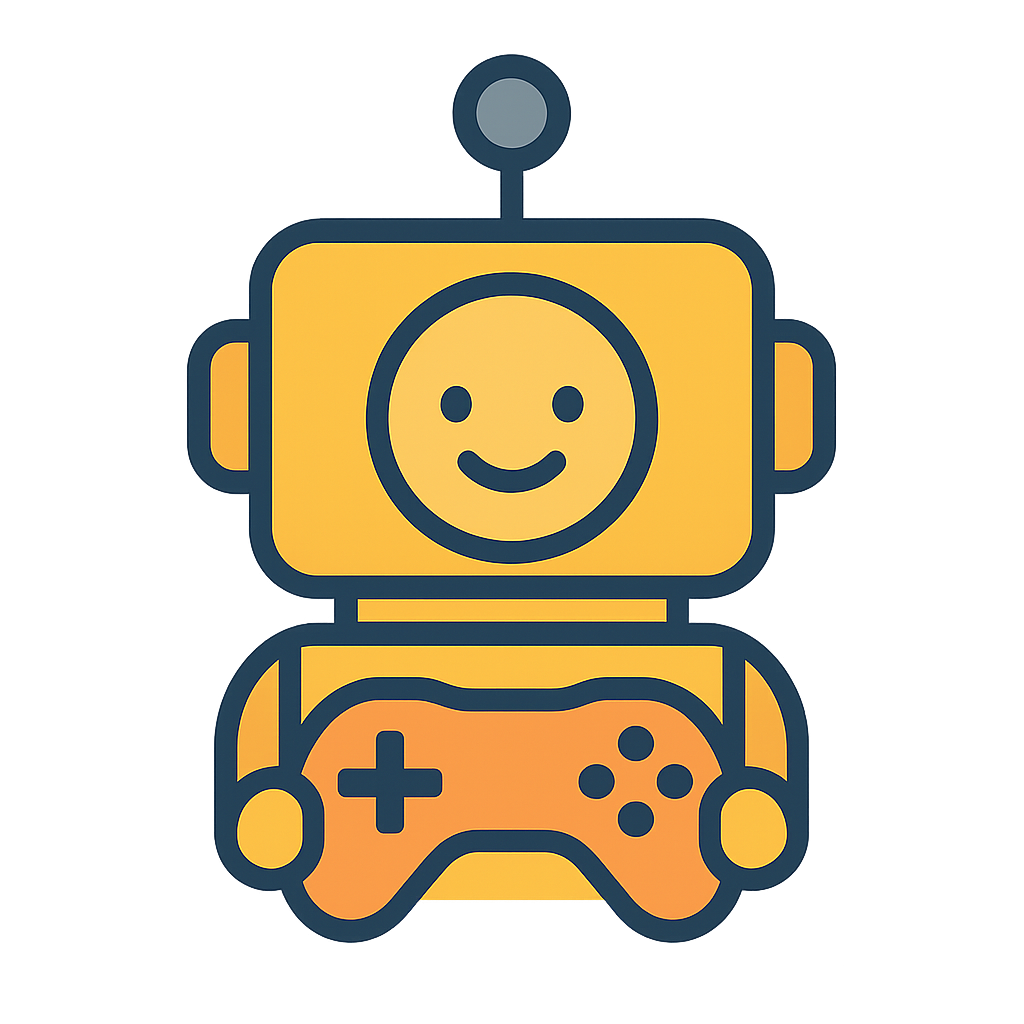}}}

\newcommand{\smallpaperlogo}{\raisebox{-.2em}{\includegraphics[height=2em]{paper-logo.png}}}

\title{
  \begin{center}
    \begin{minipage}[c]{0.12\textwidth}
      \centering
      \paperlogo
    \end{minipage}%
    \hspace{1em}
    \begin{minipage}[c]{0.80\textwidth}
      \centering
      \LARGE Leveraging LLM Agents for Automated Video Game Testing
    \end{minipage}
  \end{center}
}

\newif\ifuniqueAffiliation
\uniqueAffiliationtrue

\ifuniqueAffiliation

\author{
\href{https://orcid.org/0009-0002-9898-9517}{\includegraphics[scale=0.08]{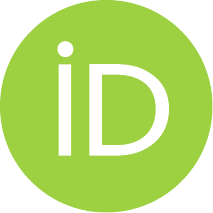}\hspace{1mm} Chengjia Wang}\\ Zhejiang University\\
    Hangzhou, China \\\texttt{cenjoy@zju.edu.cn} \\
    \And
    \href{https://orcid.org/0009-0001-1359-822X}{\includegraphics[scale=0.08]{orcid.pdf}\hspace{1mm} 
    Lanling Tang} \\
	NetEase Fuxi AI Lab\\
	Hangzhou, China \\
	\texttt{tanglanling@corp.netease.com} \\    
    \And
    \href{https://orcid.org/0009-0000-8815-1038}{\includegraphics[scale=0.08]{orcid.pdf}\hspace{1mm} Ming Yuan} \\
	NetEase Fuxi AI Lab\\
	Hangzhou, China \\
	\texttt{yuanming1220@163.com} \\
    \And
    \href{https://orcid.org/0000-0002-2888-4499}{\includegraphics[scale=0.08]{orcid.pdf}\hspace{1mm} Jiongchi Yu}\thanks{Corresponding Author.} \\
	Singapore Management University\\
    Singapore, Singapore \\
	\texttt{jcyu.2022@smu.edu.sg} \\
    \And
	\href{https://orcid.org/0000-0002-1288-6502}{\includegraphics[scale=0.08]{orcid.pdf}\hspace{1mm} Xiaofei Xie} \\
	Singapore Management University\\
    Singapore, Singapore \\
	\texttt{xfxie@smu.edu.sg} \\
    \And
	\href{https://orcid.org/0000-0002-1097-2044}{\includegraphics[scale=0.08]{orcid.pdf}\hspace{1mm} Jiajun Bu}\\
	Zhejiang University\\
	Hangzhou, China \\
	\texttt{bjj@zju.edu.cn} \\
}
\else
\usepackage{authblk}

\newbox{\orcid}\sbox{\orcid}{\includegraphics[scale=0.06]{orcid.pdf}} 
\author[1]{%
	\href{https://orcid.org/0000-0000-0000-0000}{\usebox{\orcid}\hspace{1mm}David S.~Hippocampus\thanks{\texttt{hippo@cs.cranberry-lemon.edu}}}%
}
\author[1,2]{%
	\href{https://orcid.org/0000-0000-0000-0000}{\usebox{\orcid}\hspace{1mm}Elias D.~Striatum\thanks{\texttt{stariate@ee.mount-sheikh.edu}}}%
}
\affil[1]{Department of Computer Science, Cranberry-Lemon University, Pittsburgh, PA 15213}
\affil[2]{Department of Electrical Engineering, Mount-Sheikh University, Santa Narimana, Levand}
\fi

\renewcommand{\shorttitle}{\smallpaperlogo{} Leveraging LLM Agents for Automated Video Game Testing}

\hypersetup{
pdftitle={A template for the arxiv style},
pdfsubject={q-bio.NC, q-bio.QM},
pdfauthor={David S.~Hippocampus, Elias D.~Striatum},
pdfkeywords={First keyword, Second keyword, More},
}

\begin{document}
\maketitle

\begin{abstract}
    Testing MMORPGs (Massively Multiplayer Online Role-Playing Games) is a critical yet labor-intensive task in game development due to their complexity and frequent updating nature. Traditional automated game testing approaches struggle to achieve high state coverage and efficiency in these rich, open-ended environments, while existing LLM-based game-playing approaches are limited to shallow reasoning ability in understanding complex game state-action spaces and long-complex tasks. To address these challenges, we propose \titan, an effective LLM-driven agent framework for intelligent MMORPG testing. \titan incorporates four key components to: (1) perceive and abstract high-dimensional game states, (2) proactively optimize and prioritize available actions, (3) enable long-horizon reasoning with action trace memory and reflective self-correction, and (4) employ LLM-based oracles to detect potential functional and logic bugs with diagnostic reports.

We implement \titan and evaluate it on two large-scale commercial MMORPGs spanning both PC and mobile platforms. In our experiments, \titan achieves significantly higher task completion rates (95\%) and bug detection performance compared to existing automated game testing approaches. An ablation study further demonstrates that each core component of \titan contributes substantially to its overall performance. Notably, \titan detects four previously unknown bugs that prior testing approaches fail to identify. We provide an in-depth discussion of these results, which offer guidance for new avenues of advancing intelligent, general-purpose testing systems. Moreover, \titan has been deployed in eight real-world game QA pipelines, underscoring its practical impact as an LLM-driven game testing framework.
\end{abstract}

\keywords{Large Language Model, Game Testing, Testing Agents}

\section{Introduction}

Ensuring the quality of online games is a monumental challenge in software engineering. Modern MMORPGs feature vast open worlds, nonlinear task actions, myriad interactive NPCs (non-player characters), and continuously evolving gameplay mechanics~\cite{claypool2006latency}. Testing such games thoroughly for bugs is costly, time-consuming, and often incomplete~\cite{albaghajati2020video,politowski2021survey}. Existing studies~\cite{businessresearchinsights2025gametesting} estimate that quality assurance (QA) can consume most of a game's development budget, with manual testers painstakingly playing through content to find glitches and verify game task logic. Moreover, popular games are updated frequently, often involving hundreds of code changes per day and, in some cases, more than three internal releases in a single day~\cite{yu2023gamerts}. Manual playtesting under such conditions is not only expensive and slow, but also inherently subjective and non-scalable. Human testers tend to follow common "happy paths"~\cite{anderson2008case,albaghajati2020video} while missing edge-case behaviors, and their effectiveness further declines when they are tasked with repetitive or long-duration scenarios.

Scripted test automation~\cite{alegroth2016maintenance} simulates player actions through predefined rules but is brittle and costly to maintain, as scripts must be manually crafted for each task and frequently break under game updates. More adaptive and automated DRL-based agents (e.g., Wuji~\cite{zheng2019wuji}) can learn strategies via exploration, yet they require extensive training, carefully designed rewards, and substantial computational resources, making them impractical for dynamic MMORPGs. Furthermore, DRL agents often overfit to short-term rewards and lack semantic understanding of the game tasks or items, limiting their ability to handle long-horizon objectives or detect subtle logic bugs.

The rise of large language models (LLMs) has enabled agents that combine reasoning with actions in interactive environments, as demonstrated by frameworks like LangChain~\cite{Chase_LangChain_2022} and ReAct~\cite{yao2023react}. While LLM agents show promising zero-shot generalization in text-based games and simple tasks~\cite{jeurissen2024playing,li2025frog}, directly applying them to MMORPGs remains challenging~\cite{xu2024survey}. \ding{182} MMORPGs present vast, partially observable states—covering world dynamics, NPCs, and player statistics, which is easy to exceed LLMs' perception limits due to token and memory constraints. \ding{183} Their enormous action space (movement, skills, items, dialogues) makes unguided decisions prone to invalid or irrelevant actions. Existing LLM agents~\cite{jin2024automatic} are either domain-specific (e.g., Voyager~\cite{wang2023voyager} for Minecraft) or overly generic, struggling with task logic, long-term progress, and subtle bug detection~\cite{kwa2025measuring}. Consequently, they cannot yet support scalable, efficient testing of industrial-scale MMORPGs.

\textbf{Motivation.} In real-world practice, MMORPGs are becoming increasingly larger and more complex, which makes thorough testing of sophisticated tasks even more critical. At the same time, automation is urgently needed to compensate for limited human resources. Current approaches often fail to achieve even correct task execution, let alone sufficient bug coverage and performance evaluation, thereby posing significant challenges for reliable game testing. To bridge this gap, we aim to develop an intelligent game-testing agent that addresses the above limitations. Our intuition is designing an agent that requires no task-specific training, generalizes across different games, and can complete long task sequences within complex game tasks. At the same time, it should balance task completion with maximizing game testing coverage and convenience, while effectively discovering various types of bugs, particularly functionality bugs and logic bugs. To this end, we propose \titan, an LLM-driven agent framework specifically tailored for automated MMORPG testing. \titan leverages the powerful reasoning capabilities of foundation LLMs while augmenting them with domain-specific expert knowledge to optimize the testing workflow. It provides enhanced perception, document-based extraction of game information, action recommendation, memory-based reflective reasoning, and human-in-the-loop, LLM-assisted oracles for functionality-targeted bug detection.

We implement a prototype of \titan and evaluate its performance on two large-scale commercial MMORPGs from our cooperating company. These games span both PC and mobile platforms, allowing us to assess the generalization ability and effectiveness of \titan. We construct a benchmark of 20 tasks of different difficulty levels from the two games, ranging from simple tasks with only a few steps to highly complex tasks (i.e., with over 20 steps), to thoroughly evaluate the performance of existing game-testing methods. We choose baseline approaches that include the DRL-based approach Wuji~\cite{zheng2019wuji}, and the vanilla LLM-agent-based ReAct~\cite{yao2023react} for evaluating their performance in task completion and bug detection. In addition, we conduct the same set of testing tasks with three professional internal game testers for cross-comparison.

We design a comprehensive experimental study to answer the following three key research questions:

\begin{itemize}
\item \textbf{RQ1:} Can \titan complete typical MMORPG testing tasks more effectively and efficiently than existing methods?
\item \textbf{RQ2:} How does \titan perform in detecting MMORPG bugs compared to existing approaches?
\item \textbf{RQ3:} What is the contribution of each core component of \titan to its overall performance?
\end{itemize}

Our comprehensive evaluation results demonstrate that \titan achieves significantly higher task completion rates (95\%) and bug detection performance (15) compared to automated game testing approaches (best at 82\% and 9, respectively). Among all the components, we find that the Reflective Reasoning Module contributes most to the performance of \titan, while all components are indispensable to the overall framework. Notably, TITAN detects bugs that are not identifiable by previous testing approaches and successfully uncovered four new bugs during our study, including new bug types such as Model Logic Bug and Hang Interaction Bug. \titan has been deployed in 8 real-world game QA pipelines, demonstrating the practical impact of LLM-driven game testing.
 
In summary, this paper makes the following contributions:

\begin{itemize}
    \item We design the first LLM-driven testing framework specifically for testing MMORPGs, combining the flexibility and zero-shot performance of LLMs with purpose-built components, including state abstraction, action optimization, reflective thinking, and novel oracles, to address the unique challenges of large open-world game environments. Unlike prior approaches, \titan can perceive complex game states and maintain long-horizon plans, closely mimicking how human experts strategize through game tasks.
    \item We implement \titan and evaluate it on two large-scale commercial MMORPGs on both PC and mobile platforms. We construct a diverse benchmark of 20 testing tasks across these games and, through rigorous experiments, reveal that \titan consistently outperforms baseline approaches across multiple metrics.  
    \item We provide an in-depth discussion of MMORPG testing, offering practical guidance for future development of comprehensive game testing and LLM-agent-based solutions. The \titan framework has already been adopted in the QA pipelines of several game studios, where it has uncovered new bugs and improved testing productivity. 
\end{itemize}

\section{Background and Related Works}
\label{background}

\begin{figure}[t]
    \centering
    \includegraphics[width=1\linewidth]{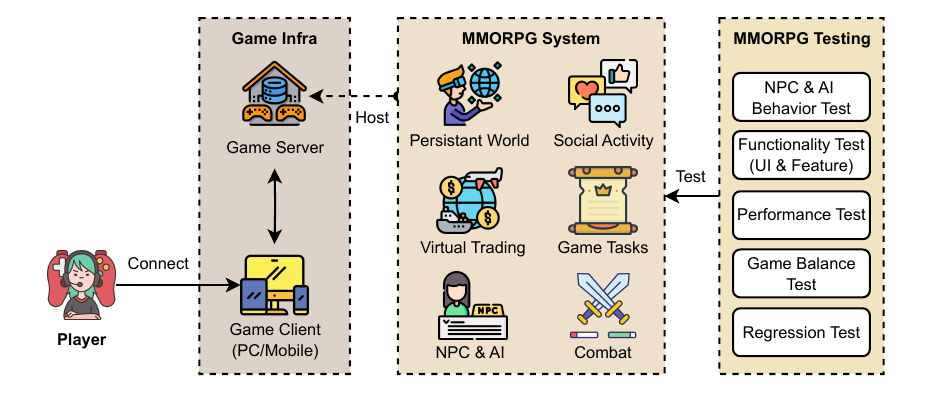}
    \caption{MMORPG Game Ecosystem and Testing.}
    \label{fig:mmorpg}
\end{figure}


In this section, we introduce the background of MMORPGs and their specific testing development. We review existing research on automated game testing and LLM-based agents for diverse applications, specifically for game exploring.

\subsection{MMORPG Testing}

Massively Multiplayer Online Role-Playing Games (MMORPGs) are large-scale online environments where players connect through game clients to persistent virtual worlds hosted on dedicated servers. As illustrated in Figure~\ref{fig:mmorpg}, MMORPG systems integrate diverse gameplay elements such as social interaction, trading, quests, and combat, all driven by complex NPC and AI behaviors. The scale and complexity of these systems introduce multifaceted testing challenges, including functionality, performance, game balance, and regression.

Compared with single-player games, massively multiplayer online role-playing games (MMORPGs) introduce unique challenges for testing due to their persistent virtual worlds, nonlinear game task structures, and high player interactivity. Early research on software quality assurance primarily focuses on traditional testing techniques such as unit testing and GUI testing~\cite{kim2007effects}. However, these methods are insufficient for addressing the dynamic and emergent gameplay patterns of MMORPGs. Subsequent studies investigate large-scale load testing and network simulation approaches to evaluate server stability and concurrency performance in MMORPG environments~\cite{cho2010scenario}. Other work contributes datasets~\cite{li2022gbgallery} and benchmarks for game testing. Further research highlights the complexity of in-game mechanics, such as non-player character (NPC) behaviors, real-time combat systems, and cooperative gameplay, which make automated testing particularly difficult~\cite{schatten2017automated,suznjevic2011mmorpg}. Player-behavior modeling has also been proposed to simulate diverse interaction patterns in MMORPGs~\cite{pfau2018towards}, aiming to expose edge cases that traditional test scripts often overlook. Despite these efforts, MMORPG testing remains costly and time-consuming, as it requires not only ensuring functional correctness but also maintaining balance, fairness, and immersion in a continuously evolving game ecosystem.

\subsection{Automated Game Testing}

Comprehensive surveys~\cite{politowski2022towards,roque2025literature,coppola2024know} highlight the breadth of issues and testing practices in the game testing domain, underscoring the immaturity of current approaches. While several techniques have been proposed for testing graphical user interface (GUI) applications~\cite{memon2007event}, these methods are often ineffective for game testing due to the complex tasks involved, which require a certain level of player intelligence. Inspired by the successful application of reinforcement learning in games and robotic navigation, recent research, such as Wuji~\cite{zheng2019wuji}, has adopted deep reinforcement learning (DRL) techniques for game testing. AdvTest~\cite{ma2024diversity} proposes a diversity-oriented testing framework that leverages constraint-guided adversarial agent training to expose more diverse failure scenarios in competitive games. Beyond adversarial training, behavior-driven development has been incorporated to guide automated testing in video games~\cite{mastain2024behavior}. Widget detection-based techniques have also been explored for industrial mobile game testing~\cite{wu2023widget,ye2021empirical}, demonstrating the potential of leveraging GUI component analysis to support automated game testing. Other studies~\cite{wu2020regression} propose reinforcement learning–based regression testing techniques that detect potential regression errors by analyzing behavioral differences between different versions of MMORPGs. Planning and learning have also been applied to video game regression testing, but remain limited in scalability~\cite{balyo2024automating}. However, these approaches rely heavily on task-specific training, making them suitable for fixed tasks such as regression testing of existing quests but difficult to adapt to new tasks and scenarios. In contrast, our method requires no training and can readily accommodate new game tasks introduced in version iterations.

\subsection{LLM-Based Game Testing}

With the rise of LLM systems, numerous attempts have been made to develop foundational agents for complex video games such as Minecraft, StarCraft II, and Civilization~\cite{lifshitz2023steve,vinyals2019grandmaster,qi2024civrealm}. Many systems rely on structured APIs and predefined action spaces (e.g., JARVIS-1~\cite{wang2024jarvis}), which constrain generalization across games. VPT~\cite{baker2022video} demonstrated end-to-end control from raw video by pretraining on human-labeled gameplay, but collecting such datasets is costly and difficult to scale. Similarly, SIMA~\cite{raad2024scaling} trained embodied agents across multiple 3D games via behavior cloning, yet remained limited by high data demands and poor generalization. More recent efforts aim at autonomous skill acquisition: VOYAGER~\cite{wang2023voyager} achieved continual exploration in Minecraft through self-generated curricula, while Cradle~\cite{zhang2023exploiting} leveraged multimodal LLMs to handle visually complex environments like Red Dead Redemption 2. Other works, such as UI-TARS~\cite{wu2024non} explored generalist agents that can interact with arbitrary UIs. LLM agents have also been applied to simulate social behaviors in virtual communities~\cite{park2023generative}, while multi-agent frameworks like Chimera~\cite{yu2025chimera} demonstrate the utility of collaborative agents for complex tasks requiring action automation. Although these approaches illustrate the promise of LLM-driven gameplay, they are often limited by domain-specific designs, expensive data requirements, or weak adaptability, which hinders their direct application to large-scale MMORPG testing.

\section{Design}

\begin{figure}[t]
    \centering
    \includegraphics[width=1\linewidth]{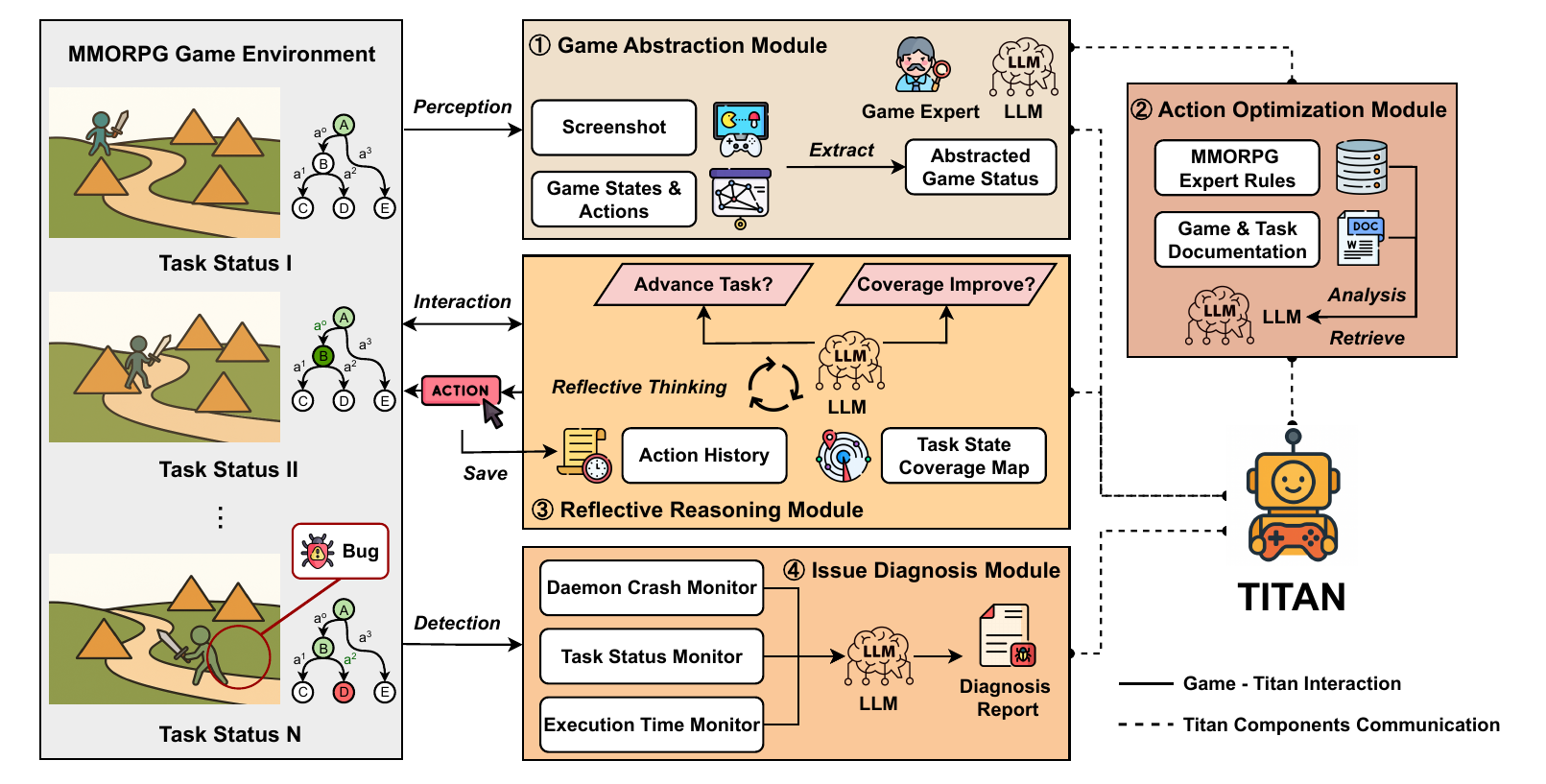}
    \caption{Overview of \titan Game Testing Framework.}
    \label{fig:overview}
\end{figure}

The \titan agent operates in a loop, interacting with the MMORPG under test, supported by four major components. In this section, we present the detailed design of each component of \titan, and their collaboration workflow is illustrated in Figure~\ref{fig:overview}.

\subsection{Overview}

We design the workflow of \titan by mirroring how expert testers operate the MMORPG testing: perceive the game state, choose meaningful actions, reflect on progress, and diagnose issues. At its core, a foundation model drives high-level reasoning, while supporting modules provide perception, action scaffolding, and diagnostic oracles for closed-loop interaction. The detailed workflow for \titan to test one task in the game is shown in Algorithm~\ref{alg:titan}.

In each loop of \titan, the perception module captures multimodal game states, which are abstracted into expert-recommended features. Game documentation and heuristics are retrieved via a RAG (Retrieval-Augmented Generation) database to guide decisions. Based on the current state and available actions, \titan selects an action, executes it, and maintains coverage maps to balance task completion with exploration. When progress stalls, reflective reasoning revisits screenshots and histories to adjust strategies. During the execution, the issue diagnosis module monitors for failures, including unfinishable tasks, crashes, and slowdowns, are logs with contextual traces. These records feed into automated reports that assist developers in bug triage. Overall, \titan adapts testing strategies dynamically, emulating the reasoning process of skilled human testers.

\subsection{Perception Abstraction Module}

The \textbf{Perception Abstraction Module} addresses the challenge of high-dimensional and heterogeneous game states in MMORPGs. Based on the expertise of experienced MMORPG testers, we observe that while different games implement their states in distinct ways, they share a set of fundamental elements and common properties (e.g., player position, health, and inventory). However, directly feeding the entire raw state into automated game testing is infeasible.

In MMORPGs, the raw state may include continuous values (e.g., map coordinates, health points, mana, timers), categorical values (e.g., current task step identifiers, weather conditions), and sets of objects (e.g., nearby NPCs, items in inventory). Passing all of this raw data to an LLM would be problematic: the input could exceed the model's context length, and the LLM may fail to interpret raw numbers correctly. To mitigate this, \titan first conducts feature selection and discretization.

We work with professional testers and designers to identify the key state factors relevant to general progress in MMORPGs, which serve as template references. These include: player location, current game objectives (from the game log), player vitals (e.g., health, mana), nearby interactable NPCs or items, and active status effects. Irrelevant data (e.g., other players' information in an online environment) is filtered out unless directly relevant to the test.

For continuous variables that are difficult to abstract, we define meaningful buckets. For instance, player health or mana is abstracted as {High, Medium, Low} with thresholds defined by experts (e.g., Low HP < 20\%). Temporal information is reduced to coarse categories (e.g., day/night) when appropriate. The abstractor then cleans and encodes the selected features into a concise textual or symbolic format. Grounding the state in high-level concepts dramatically reduces the search space. Based on predefined rules, the LLM proactively analyzes and interprets these abstract states before making decisions. This abstraction is crucial for generalization: as long as states in new games can be mapped to comparable descriptors (e.g., HP high/medium/low, semantic regions), the same prompting logic applies across different MMORPGs. To facilitate portability, abstraction rules are implemented as game-agnostic templates, configured via a metadata file specific to each game. This file defines the state units and their abstraction schemes. 

\subsection{Action Optimization Module}

The second component of the \titan framework is the \textbf{Action Optimization Module}, which tackles the decision paralysis an agent can face in open-world games. In real MMORPG testing scenarios, a player (or agent) has a vast number of possible actions at any given state, such as \textit{moving in any direction}, \textit{interacting with objects}, \textit{using skills}, and more. Most of these actions are irrelevant to the current task context or even invalid. For example, if the player is conversing with an NPC, the action \textit{Attack} is both irrelevant and inappropriate; similarly, Use of \textit{HealingPotion} makes little sense if the player's health bar is full. A naive agent without domain knowledge may still consider these options, wasting time and confusing the LLM.

We address this by defining a set of high-level action templates for the game under test. Common templates such as Move(to=Location), Talk(to=NPC), Attack(target), Use(Item), PickUp(Item), and Explore(Direction). These templates act as "tools" the LLM agent can invoke, and each can be instantiated with parameters such as NPC name, item name, or coordinates. 

Our intuitive design of \textbf{Action Optimization Module} is that we expect it to function as a pre-selection filter and advisor for the LLM's action choices. Specifically: \ding{182} Drawing on MMORPG experts' domain knowledge, we label action–state relevance rules. When a game state matches a predefined rule, the RAG database retrieves the corresponding regulation and assembles an action bundle to recommend to the LLM. \ding{183} Based on game documentation and task descriptions, the LLM also extracts action–state context pairs to generate additional recommendations. The summarized rules include common MMORPG characterizations:
\begin{itemize}
    \item \textit{1. If there is an active game objective involving a specific NPC or location, prioritize actions that involve traveling to or interacting with that NPC/location.}
    \item \textit{2. If the player's HP is low during combat, suggest using a healing item or retreating rather than ignoring survival.}
    \item \textit{3. If the inventory contains game task-related items (e.g., a key) and the objective is to use it, include an action to use the item.}
\end{itemize}

These recommendations are encoded partly as deterministic logic and partly as LLM-based heuristics. For the latter, we leverage few-shot prompting to let the agent proceed with the action prioritization (e.g., \textit{"Given the task description and the list of possible actions, output which actions are most relevant"}). This is especially useful for subtle contexts where rules alone are insufficient. For example, if a task specifies \textit{"find the hidden cave"}, the LLM can infer that \textit{Explore(north)} is more relevant than \textit{Talk(to=TownGuard)}, even if both are technically valid moves.

The optimization module then produces a recommended bundle (around five actions), which is passed to the subsequent decision step. The LLM is not strictly limited to these recommendations, as we preserve flexibility in case none are optimal, but the prompt biases the model toward these actions. A validation script is prepared to ensure that any chosen action, recommended or not, is syntactically valid before execution.

By constraining choices to a small set of contextually reasonable actions, the LLM does not waste tokens enumerating irrelevant options, and the likelihood of nonsensical decisions is minimized. This design mirrors human-like intuition filtering, much like how seasoned testers instinctively focus on the most promising actions in a given situation.

\subsection{Reflective Reasoning Module}

The \textbf{Reflective Reasoning Module} is central to the design of \titan to handle long and complex task sequences and to self-correct when facing difficulties. Given the state information obtained by the perception module, the action recommendations produced by the action optimization module, the accumulated action trace history, and an internally maintained state coverage map, the module selects the next action and interacts with the game. After each execution step, \titan appends the chosen action to the trace history and updates the coverage map with the current state. This global, task-centric memory helps the agent balance two objectives: completing the current task and exploring under-tested regions to improve coverage.

Inspired by meta-cognitive strategies used by expert testers, the module provides two key capabilities: progress monitoring and self-reflection with multimodal context, which together address the lack of timely action feedback and weak global situational awareness in traditional automated testing.

\textbf{Progress Monitoring.} As \titan executes actions, it tracks testing progress. We employ simple task-status metrics such as the percentage of key task states completed, whether the game state has advanced, and whether coverage diversity (for example, action richness) has increased. If a configured number of consecutive actions or a fixed time interval passes without measurable progress, \titan triggers the reflective reasoning mechanism so that the agent re-analyzes the situation and considers alternative strategies to break out of the current impasse.\footnote{Thresholds are configurable according to the testing budget. In our experiments, we set the trigger to 20 consecutive actions without progress based on empirical experiences.} This mechanism also helps detect logical bugs such as unreachable states or missing triggers.

\textbf{Reflection Prompting.} When reflection is triggered, \titan pauses normal decision-making and issues a reflective prompt to the LLM. The prompt includes the current abstract state, salient game information, and a concise summary of the action and state history for the ongoing task. This history enables the LLM to diagnose possible mistakes in previous steps and to surface alternative paths that may have been missed. The LLM then produces an analysis and a plan. If the reflection proposes an alternate strategy, \titan incorporates the suggested actions into subsequent steps. If it strongly indicates a bug, \titan flags the scenario as a potential defect, while optionally attempting a small number of confirming steps before finalizing the report. The exampled prompt for failing the proceeding of the task status is shown below.

\begin{tcolorbox}[colback=black!2, colframe=black!80, arc=5pt, boxrule=1.5pt, title={Reflective Reasoning Prompt}]

\textbf{[Interactive Object]}: The target NPC is nearby, but interaction attempts have failed. Navigate to the NPC's coordinates first, then try interaction.

\medskip

\textbf{[Task Progress]}: If the UI is unknown or a task overlay is open, try closing all overlays or tapping on the blank area to restore visibility.

\medskip

\textbf{[Player]}: (1) If pathfinding succeeded or interaction failed, try using the task-link action or initiating dialogue.
(2) Compare current player coordinates with task-specified coordinates; if they differ, navigate to the task coordinates first.

\medskip

\textbf{[Task Requirement]}: If the task link shows "ShowNpcGift", it indicates that after entering dialogue with the NPC, giving the gift is required.

\medskip

\textbf{[Game Information]}:  
Player Coordinates: \{x, y, z\} \\
Nearby NPCs: \{NPC1, NPC2\} \\
Game Objects: \{item1, item2\} \\
Action History: \{Action1, Action2\} \\
Task Coverage Map: \{Past Key Task1\} \\
Interactive Button Available: True

\medskip

\textbf{--- Reflection Questions ---}

1. What are possible causes for no progress?   \\
2. What concrete sequence of actions should be tried next (navigation, interaction method, interface cleanup, etc.)?   \\
3. Is this likely a bug? If yes, what type (logic/hang/interaction), and what evidence supports that (coordinate mismatch, action failures, UI overlay, etc.)?

\medskip

\textbf{[Constraints]}:  
- Suggest concrete actions: e.g., navigate to NPC, click/tap interaction button, give gift, close UI overlay.\\  
- Prioritize actions that explore new or untested game states or paths.\\  
- If multiple plans are possible, order them by plausibility.

\end{tcolorbox}

\textbf{Execution Coverage Memory.} Beyond within-episode reflection, \titan reuses knowledge from prior episodes. We maintain a persistent memory of abstract states and actions explored in earlier runs of the same or similar game tasks, tagging transitions with outcomes such as success or bug encountered. This induces a coverage map represented as a state–action transition graph. During new tests, \titan consults this memory to inform action selection. For example, if a particular state–action pair has been tried many times without yielding progress, the agent deprioritizes it in favor of untried or promising alternatives. This cross-run reflectiveness resembles how human testers learn from prior sessions and helps systematically increase coverage across runs without unguided exploration.

Together, these mechanisms ensure that \titan does not naively step through actions. Instead, it continuously reasons about progress, adapts its strategy when needed, and accumulates experience to become more effective over time.

\subsection{Issue Diagnosis Module}


During execution, the \textbf{Issue Diagnosis Module} of \titan encapsulates our approach to automated detection and reporting of potential MMORPG issues. In testing complex systems such as large-scale games, some oracles are straightforward (e.g., a crash is always a bug), but many others require contextual, functional, or logical reasoning, which has long been a major challenge. To address this, \titan integrates several complementary oracles:

\textbf{Crash Monitor}. This component listens for unhandled exceptions, crashes, or forced terminations of the game during testing. We instrument the game client to capture crash events. Any crash is immediately recorded as a bug, along with the state and action trace that triggered it.

\textbf{Task Status Monitor}. If the \textbf{Reflective Reasoning Module} exceeds its threshold for stalled task execution, \titan concludes that the current task is likely not completable under existing conditions. Instead of silently failing, the module triggers an oracle that flags a potential stuck issue. The LLM's reflective output is used to generate a structured diagnosis report that includes: \ding{182} a brief description of the tasks and its goal, \ding{183} a summary of the agent's attempted actions, \ding{184} analysis of why \titan believes the game task is stuck, and \ding{185} supporting evidence such as abstract state details or screenshots. These reports are saved for human review and can also be submitted as tickets for developers. Depending on staffing constraints and iteration urgency, the acceptance threshold for these reports may be adjusted so that all flagged cases can serve as potential issues for further triage.

\textbf{Execution Time Monitor}. Inspired by established research in software engineering, which considers differential testing~\cite{song2025multi,castellano2022explaining}, we design an oracle that monitors execution time for both actions and tasks. Each action type has an expected duration based on the average execution time recorded (This oracle will be disabled on the first time of execution). If \titan observes a significant deviation (e.g., an action that typically completes within one second now takes more than 10 seconds, or a game task expected to finish in five minutes continues for 15 minutes without progress), it will flag a performance anomaly. Such anomalies may indicate issues such as infinite loops, server lag, or resource leaks. These problems do not necessarily crash the game but can severely degrade user experience and are often overlooked by other oracles. The sensitivity of this oracle can be tuned: a narrow deviation threshold yields higher detection coverage but increases false positives, whereas a wider threshold reduces false positives at the risk of missing subtle bugs. 

By combining these assembled oracles, \titan automatically detects a broad spectrum of issues. The overall false alarm rate of \titan is 30\%, which is relatively low in practice, with most false positives caused by imperfect environment adaptation during state abstraction. 
The synergy between the agent’s exploratory behavior and these detection mechanisms produces a robust and autonomous framework for MMORPG testing.

\begin{algorithm}[t]
\caption{Algorithm of \titan framework.}
\label{alg:titan}
\DontPrintSemicolon
\SetKwInOut{Input}{Input}\SetKwInOut{Output}{Output}
\Input{
  Game environment $\mathcal{G}$ (control APIs, state info, screenshots);\\
  Expert knowledge $\mathcal{K}$; Reflection \& oracle thresholds $(X,Y,T)$;\\
  Action feasibility rules $\Phi$, 
  Test objective $\mathcal{O}$ (e.g., complete task $q$, maximize coverage).
}
\Output{
  Potential issue set $\mathbb{I}$; Diagnosis reports $\mathbb{R}$.
}
\BlankLine

\textbf{Initialize:} $\mathbb{I}\leftarrow\varnothing$, $\mathbb{R}\leftarrow\varnothing$;\ 
Global coverage map $\mathbb{C}\leftarrow$ load prior abstract state map;\ 
Action history $\mathbb{H}\leftarrow\varnothing$.\;

\While{scenario not terminated}{
  \tcp{1) Perception \& Abstraction}
  Raw state $s_{\text{raw}}\leftarrow$ \textsc{ReadState}$(\mathcal{G})$;\ 
  Screenshot $I \leftarrow$ \textsc{Capture}$(\mathcal{G})$ (in reflective mode);\;
  $s \leftarrow$ \textsc{StateAbstractor}$(s_{\text{raw}}, \mathcal{K})$;\;
  \textsc{UpdateCoverage}$(\mathbb{C}, s)$;\;
  
  \tcp{2) Action Space Optimization}
  $\mathcal{A}_{\text{cand}}\leftarrow$ \textsc{EnumerateTemplates}$(\mathcal{A}, s)$; $\mathcal{A} \leftarrow \textsc{ReadActions}(\mathcal{G})$;\;
  $\mathcal{A}_{\text{feas}}\leftarrow \{a\in\mathcal{A}_{\text{cand}} \mid \Phi(a,s)=\textsf{true}\}$;\;
  $\mathcal{A}_{\text{rec}}\leftarrow$ \textsc{LLM-Recommend}$(s,\mathcal{O},\mathcal{A}_{\text{feas}},\mathbb{C})$;\;

  \tcp{3) LLM Reasoning \& Act}
  $a^\star\leftarrow$ \textsc{LLM-Decide}$(s,\mathcal{O},\mathcal{A}_{\text{rec}},\mathbb{H})$;\;
  \If{\textsc{Validate}$(a^\star,\Phi)=\textsf{false}$}{
    $a^\star\leftarrow$ \textsc{Fallback}$(\mathcal{A}_{\text{rec}})$\;
  }
  $o \leftarrow$ \textsc{Execute}$(\mathcal{G}, a^\star)$;\ 
  $\mathbb{H}\leftarrow\mathbb{H}\cup\{(s,a^\star,o)\}$;\;

  \tcp{4) Execution Monitoring \& Reflection}
  \If{\textsc{NoCoverageImprove}$(\mathbb{H},X)$ \textbf{or} \textsc{Timeout}$(T)$}{
    $\pi_{\text{refl}}\leftarrow$ \textsc{LLM-Reflect}$(s,\mathbb{H},I,\mathcal{O})$;\;
    \If{\textsc{BugSuspected}$(\pi_{\text{refl}})$}{
      $(\iota,\rho)\leftarrow$ \textsc{MakeDiagnosis}$(\pi_{\text{refl}},s,\mathbb{H},I)$;\ 
      $\mathbb{I}\leftarrow\mathbb{I}\cup\{\iota\}$,\ \ $\mathbb{R}\leftarrow\mathbb{R}\cup\{\rho\}$\;
      \If{\textsc{EscalationCount}$\ge Y$}{\textbf{break}} 
    }
    \Else{
      \textsc{InjectPlan}$(\mathcal{A}_{\text{rec}},\pi_{\text{refl}})$; \textbf{continue}\;
    }
  }

  \tcp{5) Optimized Oracle for MMORPG}
  \If{\textsc{CrashDetected}$(\mathcal{G})$}{
    $(\iota,\rho)\leftarrow$ \textsc{ReportCrash}$(s,\mathbb{H})$;\
    $\mathbb{I}\leftarrow\mathbb{I}\cup\{\iota\}$,\ $\mathbb{R}\leftarrow\mathbb{R}\cup\{\rho\}$;\ \textbf{break}\;
  }
  \If{\textsc{LogicAnomaly}$(\mathbb{H})$ \textsc{or} \textsc{TimeAnomaly}$(\mathbb{H})$} {
    $(\iota,\rho)\leftarrow$ \textsc{ReportToExpert}$(s,\mathbb{H},\mathcal{K})$;\
    $\mathbb{I}\leftarrow\mathbb{I}\cup\{\iota\}$,\ $\mathbb{R}\leftarrow\mathbb{R}\cup\{\rho\}$\;
  }
  \If{\textsc{TaskComplete}$(\mathcal{G},\mathcal{O})$}{\textbf{break}}
}
\Return $(\mathbb{I},\mathbb{R})$
\end{algorithm}

\section{Experiment Setup}

\subsection{Target Games}
We select two large commercial MMORPGs from our collaborating companies for evaluation, referred to as Game A and Game B for anonymity. Specifically, Game A is a PC-based 3D MMORPG that has been released for over five years with an open-world martial arts theme. It features real-time combat, an extensive gaming system, and large world areas, with a daily active player base in the hundreds of thousands. Game B is a mobile MMORPG (Android/iOS) that has been released for ten years and has millions of downloads. It adopts a top-down perspective, tap-based controls, and auto-pathfinding to support diverse game tasks.

These two games differ in thematic style, control schemes, and player scale, together providing a robust testbed for assessing \titan's generalization ability. Importantly, we did not fine-tune \titan's foundation model backbone, while only the state abstraction templates were adapted to each game's API and content. This adaptation requires a one-time setup effort of less than five minutes per game.

\subsection{Evaluated Tasks}

For the two selected games, we categorize all tasks according to their difficulty and complexity, based on expected completion time and overall state–action complexity. Tasks are grouped into three categories: \textbf{Simple} (approximately 10 state-action pairs), \textbf{Normal} (around 20 state-action pairs), and \textbf{Hard} (over 20 state-action pairs). Due to limited experimental time, we curated ten tasks from each game, yielding a total of twenty tasks, evenly sampled across the three categories, as summarized in Table~\ref{tab:task-list}.

Since many tasks correspond to multi-step task lines, some scenarios involve reaching hard-to-access locations or triggering optional game task branches. The bugs used in our evaluation are previously reported issues from the game developers, drawn from six months of bug tracing and reproducibility analysis. Importantly, neither TITAN nor the baseline methods are informed of the presence or location of the bugs; they are only provided with the task description and required to play through the scenario, ideally discovering the bug during execution.


\begin{table}[t]
\centering
\caption{Selected Task List}
\label{tab:task-list}
\resizebox{.6\textwidth}{!}{
\begin{tabular}{c|c|cc|cc}
\hline
\rowcolor[HTML]{F2F1F1} 
\cellcolor[HTML]{F2F1F1}                                & \cellcolor[HTML]{F2F1F1}                                      & \multicolumn{2}{c|}{\cellcolor[HTML]{F2F1F1}\textbf{Game A}} & \multicolumn{2}{c}{\cellcolor[HTML]{F2F1F1}\textbf{Game B}} \\ \cline{3-6} 
\rowcolor[HTML]{F2F1F1} 
\multirow{-2}{*}{\cellcolor[HTML]{F2F1F1}\textbf{Task}} & \multirow{-2}{*}{\cellcolor[HTML]{F2F1F1}\textbf{Difficutly}} & \textbf{State}                & \textbf{Action}                & \textbf{State}                & \textbf{Action}               \\ \hline
1                                                       & Simple                                                        & 9                             & 6                            & 8                             & 5                           \\
2                                                       & Simple                                                        & 11                            & 8                            & 6                             & 5                           \\
3                                                       & Simple                                                        & 15                            & 8                            & 13                            & 9                           \\
4                                                       & Normal                                                        & 14                            & 11                           & 19                            & 11                          \\
5                                                       & Normal                                                        & 19                            & 11                           & 15                            & 14                          \\
6                                                       & Normal                                                        & 20                            & 13                           & 16                            & 14                          \\
7                                                       & Normal                                                        & 26                            & 18                           & 18                            & 18                          \\
8                                                       & Hard                                                          & 23                            & 21                           & 28                            & 20                          \\
9                                                       & Hard                                                          & 32                            & 25                           & 25                            & 23                          \\
10                                                      & Hard                                                          & 33                            & 27                           & 25                            & 30                          \\ \hline
\end{tabular}
}
\end{table}

\subsection{Baseline Methods}
We compare the performance of \titan in terms of efficiency and effectiveness against three baseline approaches: 

\textbf{DRL-based Approach.} We select Wuji~\cite{zheng2019wuji} as the state-of-the-art DRL-based method for game testing. Specifically, we follow the configuration and design described in the original paper and re-implement Wuji's evolutionary DRL algorithm with two objectives: \ding{182} task completion, with a reward of +1 for finishing the task and partial rewards for intermediate steps when available, and \ding{183} exploration, with rewards for visiting novel state features, consistent with the original design. We train separate agents for Game A and Game B under on-the-fly testing settings. For each evaluated task, the Wuji agent continues training and exploration until it either discovers the bug, completes the task, or reaches the time limit of 60 minutes.

\textbf{LLM Agent-Based Approach.} We select one of the most widely used LLM-based agent frameworks ReAct~\cite{yao2023react} for comparison. We employ the same foundation LLM models as TITAN, which are detailed in ~\ref{sec:foundation_models} and are prompted to alternately observe the game state and output reasoning steps and corresponding actions. The agent is provided with the raw state dump of the game and is free to output any action; we attempt to execute the action if it is syntactically valid. This baseline is designed to evaluate how an off-the-shelf LLM agent performs on MMORPG testing tasks without additional domain-specific optimization.

\textbf{Human Testing.} We include professional testers from our collaborating company as another baseline reference. We recruit three experienced MMORPG QA testers who are familiar with general game testing practices but new to these two specific games. Each tester plays through a subset of the selected tasks three times, and we compute the average performance to reduce randomness. Testers are provided with the task descriptions and game manuals and are allowed to manage their own time and notes as they see fit. For each task, we measure task completion, test coverage, and whether the tester successfully detected the bug.

Although we considered comparing against purely scripted testing, such baselines are impractical due to the diverse states and flexible scenarios in the selected tasks, particularly in the complex category, where scripting unique solutions is infeasible. Instead, the human results can be viewed as a proxy for well-designed manual test cases, albeit with lower testing efficiency.

\subsection{Foundation Models and Experiment Devices}
\label{sec:foundation_models}

During our experiments, we used OpenAI GPT-4o as the backbone model for both \titan and ReAct, due to API budget constraints. We set the temperature to 0 to ensure deterministic outputs and reduce run-to-run variance. The prompt design included a system message assigning the agent the role of a tester along with game-specific context. Few-shot examples were provided for both normal decision-making and the reflection stage. All experiments were conducted on two high-performance servers equipped with AMD EPYC 9654 96-core processors running at 2.40GHz, providing a total of 192 cores and 1 TB of RAM.

\textbf{Reliability of Experiment Results.} All results are averaged over five runs for \titan and the automated testing baselines (Wuji, ReAct). The results of Wuji are averaged over five independent training seeds. Human testing results are aggregated across the recruited testers three times to further mitigate randomness.

\section{Evaluation Result}

In this section, we present the experimental results and address each research question in turn. Overall, \titan demonstrates superior performance across all key metrics, validating the advantages of its LLM-driven yet structured approach to game testing. Compared with existing automated methods, \titan achieves higher effectiveness and robustness, while also outperforming human experts in testing efficiency.

\subsection{RQ1: Task Completion Performance}

We first examine the performance of different methods on two key dimensions: task completion ability and state exploration capacity. In practical MMORPG testing, given the increasing complexity of modern games, the ability to reliably complete game tasks forms the foundation of quality assurance. We evaluate all methods on twenty tasks drawn from the two selected games, measuring both their task success rate and the state coverage achieved across multiple runs. Table~\ref{tab:rq1} summarizes the results.

We find that \titan completes 95\% of the evaluated tasks successfully, closely approaching the 100\% success rate achieved by professional human testers. By contrast, the baselines lagged: the DRL-based Wuji completed approximately 82\% of tasks, while the vanilla LLM agent ReAct managed only 83\%. This performance gap underscores \titan's effectiveness in executing complex game tasks. Notably, \titan solved all simple and normal tasks and failed only twice on complex tasks involving very long action chains. In comparison, ReAct often became stuck early in the task, and Wuji occasionally failed to converge to a viable strategy within the allowed time limit.

Beyond task success, we also measured the state coverage achieved by each method. \titan consistently visited the highest number of unique abstract states, covering on average 73.26\% of the combined unique state space across all scenarios. Wuji, despite its exploratory design, reached about 54.54\% coverage, while ReAct achieved only 59.98\%, often becoming trapped in repetitive patterns due to insufficient guidance. The advantage of \titan lies in its reflective reasoning and coverage memory, which encourage exploration of alternative paths in repeated runs. Wuji's exploration, though broad, remained constrained by policy biases learned during training.

These results suggest that \titan not only excels at completing core tasks but also achieves substantially higher coverage, which is critical for thorough testing. Effective MMORPG testing requires balancing task completion with exploration of untested states to uncover subtle or edge-case bugs. \titan demonstrates this balance by pursuing main objectives while strategically deviating to explore uncharted scenarios.

\begin{tcolorbox}[size=title,opacityfill=0.1]
\noindent
\textbf{Finding 1:} \titan not only achieves a superior task completion rate (95\%) compared to Wuji (82\%) and the ReAct (83\%), but also exhibits markedly higher state exploration capacity, visiting the most unique abstract states among all methods. This demonstrates that \titan strikes an effective balance between reliably finishing game tasks and exploring untested game state spaces.
\end{tcolorbox}

\begin{table}[t]
\centering
\caption{Comparison result of task completion success rates (SR) and state coverage (CV) of two MMORPGs.}
\label{tab:rq1}
\resizebox{.8\textwidth}{!}{
\begin{tabular}{ccccccccc}
\hline
\multicolumn{1}{c|}{}                                & \multicolumn{8}{c}{\textbf{Game A}}                                                                                                                                                                                                                                             \\ \cline{2-9} 
\multicolumn{1}{c|}{}                                & \multicolumn{2}{c|}{\textbf{TITAN}}                                       & \multicolumn{2}{c|}{\textbf{ReAct}}                                       & \multicolumn{2}{c|}{\textbf{Wuji}}                                        & \multicolumn{2}{c}{\textbf{Human}}          \\ \cline{2-9} 
\multicolumn{1}{c|}{\multirow{-3}{*}{\textbf{Task}}} & SR (\%)              & \multicolumn{1}{c|}{CV (\%)}                       & SR (\%)              & \multicolumn{1}{c|}{CV (\%)}                       & SR (\%)              & \multicolumn{1}{c|}{CV (\%)}                       & SR (\%)              & CV (\%)              \\ \hline
\multicolumn{1}{c|}{1}                               & 100                  & \multicolumn{1}{c|}{82.11}                         & 60                   & \multicolumn{1}{c|}{78.95}                         & 60                   & \multicolumn{1}{c|}{68.42}                         & 100                  & 15.79                \\
\multicolumn{1}{c|}{2}                               & 100                  & \multicolumn{1}{c|}{76.25}                         & 100                  & \multicolumn{1}{c|}{33.75}                         & 100                  & \multicolumn{1}{c|}{100.00}                        & 100                  & 29.67                \\
\multicolumn{1}{c|}{3}                               & 100                  & \multicolumn{1}{c|}{50}                            & 100                  & \multicolumn{1}{c|}{24}                            & 100                  & \multicolumn{1}{c|}{64.00}                         & 100                  & 40                   \\
\multicolumn{1}{c|}{4}                               & 100                  & \multicolumn{1}{c|}{57.26}                         & 100                  & \multicolumn{1}{c|}{48.38}                         & 80                   & \multicolumn{1}{c|}{66.13}                         & 100                  & 27.42                \\
\multicolumn{1}{c|}{5}                               & 100                  & \multicolumn{1}{c|}{93.43}                         & 100                  & \multicolumn{1}{c|}{27.27}                         & 80                   & \multicolumn{1}{c|}{42.42}                         & 100                  & 17.17                \\
\multicolumn{1}{c|}{6}                               & 100                  & \multicolumn{1}{c|}{64.5}                          & 100                  & \multicolumn{1}{c|}{62}                            & 100                  & \multicolumn{1}{c|}{68.87}                         & 100                  & 21.19                \\
\multicolumn{1}{c|}{7}                               & 100                  & \multicolumn{1}{c|}{68.75}                         & 100                  & \multicolumn{1}{c|}{40.63}                         & 80                   & \multicolumn{1}{c|}{25.62}                         & 100                  & 28.13                \\
\multicolumn{1}{c|}{8}                               & 80                   & \multicolumn{1}{c|}{68.61}                         & 80                   & \multicolumn{1}{c|}{35}                            & 60                   & \multicolumn{1}{c|}{42.50}                         & 100                  & 18.33                \\
\multicolumn{1}{c|}{9}                               & 100                  & \multicolumn{1}{c|}{84.55}                         & 40                   & \multicolumn{1}{c|}{61.38}                         & 80                   & \multicolumn{1}{c|}{73.98}                         & 100                  & 29.67                \\
\multicolumn{1}{c|}{10}                              & 80                   & \multicolumn{1}{c|}{76.31}                         & 60                   & \multicolumn{1}{c|}{48.83}                         & 0                    & \multicolumn{1}{c|}{74.56}                         & 100                  & 41.22                \\ \hline
\rowcolor[HTML]{F2F0F0} 
\multicolumn{1}{c|}{\cellcolor[HTML]{F2F0F0}Avg}     & 96                   & \multicolumn{1}{c|}{\cellcolor[HTML]{F2F0F0}72.17} & 84                   & \multicolumn{1}{c|}{\cellcolor[HTML]{F2F0F0}49.82} & 74                   & \multicolumn{1}{c|}{\cellcolor[HTML]{F2F0F0}62.65} & 100                  & 26.86                \\ \hline
\multicolumn{1}{l}{}                                 & \multicolumn{1}{l}{} & \multicolumn{1}{l}{}                               & \multicolumn{1}{l}{} & \multicolumn{1}{l}{}                               & \multicolumn{1}{l}{} & \multicolumn{1}{l}{}                               & \multicolumn{1}{l}{} & \multicolumn{1}{l}{} \\ \hline
\multicolumn{1}{c|}{}                                & \multicolumn{8}{c}{\textbf{Game B}}                                                                                                                                                                                                                                             \\ \cline{2-9} 
\multicolumn{1}{c|}{}                                & \multicolumn{2}{c|}{\textbf{TITAN}}                                       & \multicolumn{2}{c|}{\textbf{ReAct}}                                       & \multicolumn{2}{c|}{\textbf{Wuji}}                                        & \multicolumn{2}{c}{\textbf{Human}}          \\ \cline{2-9} 
\multicolumn{1}{c|}{\multirow{-3}{*}{\textbf{Task}}} & SR (\%)              & \multicolumn{1}{c|}{CV (\%)}                       & SR (\%)              & \multicolumn{1}{c|}{CV (\%)}                       & SR (\%)              & \multicolumn{1}{c|}{CV (\%)}                       & SR (\%)              & CV (\%)              \\ \cline{2-9} 
\multicolumn{1}{c|}{1}                               & 100                  & \multicolumn{1}{c|}{88}                            & 80                   & \multicolumn{1}{c|}{47.61}                         & 100                  & \multicolumn{1}{c|}{65.71}                         & 100                  & 65.63                \\
\multicolumn{1}{c|}{2}                               & 100                  & \multicolumn{1}{c|}{88}                            & 100                  & \multicolumn{1}{c|}{60}                            & 100                  & \multicolumn{1}{c|}{83.3}                          & 100                  & 88.42                \\
\multicolumn{1}{c|}{3}                               & 100                  & \multicolumn{1}{c|}{81.19}                         & 100                  & \multicolumn{1}{c|}{50.8}                          & 100                  & \multicolumn{1}{c|}{80.36}                         & 100                  & 58.95                \\
\multicolumn{1}{c|}{4}                               & 100                  & \multicolumn{1}{c|}{97.37}                         & 100                  & \multicolumn{1}{c|}{81.65}                         & 80                   & \multicolumn{1}{c|}{68.37}                         & 100                  & 19.58                \\
\multicolumn{1}{c|}{5}                               & 100                  & \multicolumn{1}{c|}{87}                            & 100                  & \multicolumn{1}{c|}{71.42}                         & 80                   & \multicolumn{1}{c|}{13.81}                         & 100                  & 66.04                \\
\multicolumn{1}{c|}{6}                               & 100                  & \multicolumn{1}{c|}{70.3}                          & 80                   & \multicolumn{1}{c|}{64.89}                         & 80                   & \multicolumn{1}{c|}{34.92}                         & 100                  & 22.01                \\
\multicolumn{1}{c|}{7}                               & 80                   & \multicolumn{1}{c|}{46.7}                          & 60                   & \multicolumn{1}{c|}{75.71}                         & 100                  & \multicolumn{1}{c|}{58.97}                         & 100                  & 32.98                \\
\multicolumn{1}{c|}{8}                               & 80                   & \multicolumn{1}{c|}{73.27}                         & 80                   & \multicolumn{1}{c|}{71.66}                         & 80                   & \multicolumn{1}{c|}{54.05}                         & 100                  & 32                   \\
\multicolumn{1}{c|}{9}                               & 100                  & \multicolumn{1}{c|}{57.1}                          & 60                   & \multicolumn{1}{c|}{63.18}                         & 100                  & \multicolumn{1}{c|}{28.69}                         & 100                  & 69.87                \\
\multicolumn{1}{c|}{10}                              & 80                   & \multicolumn{1}{c|}{95.63}                         & 60                   & \multicolumn{1}{c|}{97.86}                         & 80                   & \multicolumn{1}{c|}{67.43}                         & 100                  & 23.33                \\ \hline
\rowcolor[HTML]{F2F0F0} 
\multicolumn{1}{c|}{\cellcolor[HTML]{F2F0F0}Avg}     & 94                   & \multicolumn{1}{c|}{\cellcolor[HTML]{F2F0F0}74.34} & 82                   & \multicolumn{1}{c|}{\cellcolor[HTML]{F2F0F0}70.13} & 90                   & \multicolumn{1}{c|}{\cellcolor[HTML]{F2F0F0}48.43} & 100                  & 40.88                \\ \hline
\end{tabular}
}
\end{table}

\subsection{RQ2: Bug Detection Performance}

Building on task-completion results, a more important metric is whether the agent can detect the bugs seeded in those tasks. We therefore evaluate each method's bug-detection performance and execution efficiency over five runs per task. The results are shown in Table~\ref{tab:rq2}. Across the nine bugs, \titan correctly detected and reported 82\% of them, outperforming prior detection methods and human experts. By comparison, Wuji detected about 45.5\%, ReAct detected around 45.5\%, and human testers detected approximately 18\%, while they often miss subtle issues or take too long to notice them.

We further analyze detection by bug category, as shown in Table~\ref{tab:bug-dist}:

\textbf{Crash Bugs.} All approaches that reached the crash trigger detected the failure. The primary difference was reachability. \titan and Wuji triggered all crashes in Game A. In Game B, Wuji reached only about 30\% of the crash points. The LLM baseline ReAct missed several crashes, detecting only about 80\% to 83\% overall, because it often failed to progress far enough to activate the trigger. Human testers, constrained to five attempts and focused on completing the task efficiently, found fewer crash cases even though they finished the tasks.

\textbf{Hang Bugs.} Beyond crashes, we observe corner cases where progression stalled or tasks appeared to halt. Agents are well-suited to detect such conditions. In our study, certain tasks required using an item, with two valid interaction pathways: pressing an on-screen interaction key or selecting the item from the inventory. The Wuji tend to choose a single habitual pathway, whereas LLM agents vary their interaction patterns, improving the chance of exposing the stall. \titan detects two hang bugs in Game A, and ReAct detects one.

\textbf{Logic Bugs.} These are the most difficult to uncover in daily testing. \titan's oracle design substantially improved the identification and analysis of stuck conditions. For example, one task in Game A required defeating six NPCs, yet after five defeats, the tasks would sometimes advance to the next step. \titan tries multiple combat strategies and surfaces the inconsistency as a potential logic defect. In contrast, Wuji often repeats a single auto-combat action, and human testers sometimes advance without noticing the mismatch, which leads to missed detections.

Regarding execution time, we find that \titan achieves the highest efficiency among all methods. Human testers required the longest time overall, whereas \titan explored the game space much more efficiently. For Game A, ReAct performed relatively well, handling tasks more quickly than Wuji. However, for Game B, ReAct became significantly slower, sometimes even slower than human testers. This slowdown is attributable to the design of Game B, which relies heavily on mobile-style tap controls and auto-pathfinding. These mechanics produced verbose raw state dumps that confused the LLM, leading to inefficient decision-making and longer execution times.

Overall, \titan achieved higher detection rates and did so more efficiently, aided by reflection and category-aware oracles. The combination of progress monitoring, strategy revision, and structured diagnosis enabled \titan to uncover logic and functionality errors more reliably than the baselines.

\begin{tcolorbox}[size=title,opacityfill=0.1]
\noindent
\textbf{Finding 2:} \titan substantially outperforms baseline methods in detecting seeded bugs — not merely by increasing detection rate overall, but by detecting hard‐to‐reach crash, hang, and logic bugs that baselines (Wuji, ReAct) and human testers often miss. Moreover, \titan does so with greater execution efficiency and variation in interaction style, enabling early triggering of various bugs.
\end{tcolorbox}

\begin{table}[t]
\centering
\caption{Comparison result of bug detection performance and task execution time of two MMORPGs.}
\label{tab:rq2}
\resizebox{.8\textwidth}{!}{
\begin{tabular}{ccccccccc}
\hline
\multicolumn{1}{c|}{}                                & \multicolumn{8}{c}{\textbf{Game A}}                                                                                                                                                                                                                                       \\ \cline{2-9} 
\multicolumn{1}{c|}{}                                & \multicolumn{2}{c|}{\textbf{TITAN}}                                     & \multicolumn{2}{c|}{\textbf{ReAct}}                                     & \multicolumn{2}{c|}{\textbf{Wuji}}                                      & \multicolumn{2}{c}{\textbf{Human}}          \\ \cline{2-9} 
\multicolumn{1}{c|}{\multirow{-3}{*}{\textbf{Task}}} & \#Bug                & \multicolumn{1}{c|}{Time (m)}                    & \#Bug                & \multicolumn{1}{c|}{Time (m)}                    & \#Bug                & \multicolumn{1}{c|}{Time (m)}                    & \#Bug                & Time (m)             \\ \hline
\multicolumn{1}{c|}{1}                               & 1                    & \multicolumn{1}{c|}{16}                          & 1                    & \multicolumn{1}{c|}{7}                           & 1                    & \multicolumn{1}{c|}{7}                           & 0                    & 10                   \\
\multicolumn{1}{c|}{2}                               & 1                    & \multicolumn{1}{c|}{36}                          & 1                    & \multicolumn{1}{c|}{18}                          & 0                    & \multicolumn{1}{c|}{96}                          & 0                    & 65                   \\
\multicolumn{1}{c|}{3}                               & 1                    & \multicolumn{1}{c|}{18}                          & 1                    & \multicolumn{1}{c|}{11}                          & 1                    & \multicolumn{1}{c|}{17}                          & 1                    & 46                   \\
\multicolumn{1}{c|}{4}                               & 1                    & \multicolumn{1}{c|}{42}                          & 1                    & \multicolumn{1}{c|}{20}                          & 1                    & \multicolumn{1}{c|}{52}                          & 0                    & 127                  \\
\multicolumn{1}{c|}{5}                               & 0                    & \multicolumn{1}{c|}{64}                          & 0                    & \multicolumn{1}{c|}{20}                          & 0                    & \multicolumn{1}{c|}{59}                          & 0                    & 117                  \\
\multicolumn{1}{c|}{6}                               & 0                    & \multicolumn{1}{c|}{53}                          & 0                    & \multicolumn{1}{c|}{119}                         & 0                    & \multicolumn{1}{c|}{58}                          & 0                    & 80                   \\
\multicolumn{1}{c|}{7}                               & 1                    & \multicolumn{1}{c|}{63}                          & 0                    & \multicolumn{1}{c|}{33}                          & 0                    & \multicolumn{1}{c|}{27}                          & 0                    & 138                  \\
\multicolumn{1}{c|}{8}                               & 0                    & \multicolumn{1}{c|}{148}                         & 0                    & \multicolumn{1}{c|}{146}                         & 0                    & \multicolumn{1}{c|}{114}                         & 0                    & 149                  \\
\multicolumn{1}{c|}{9}                               & 2                    & \multicolumn{1}{c|}{135}                         & 0                    & \multicolumn{1}{c|}{191}                         & 1                    & \multicolumn{1}{c|}{137}                         & 0                    & 173                  \\
\multicolumn{1}{c|}{10}                              & 2                    & \multicolumn{1}{c|}{129}                         & 1                    & \multicolumn{1}{c|}{84}                          & 1                    & \multicolumn{1}{c|}{210}                         & 1                    & 233                  \\ \hline
\rowcolor[HTML]{F2F0F0} 
\multicolumn{1}{c|}{\cellcolor[HTML]{F2F0F0}Total}   & 9                    & \multicolumn{1}{c|}{\cellcolor[HTML]{F2F0F0}704} & 5                    & \multicolumn{1}{c|}{\cellcolor[HTML]{F2F0F0}649} & 5                    & \multicolumn{1}{c|}{\cellcolor[HTML]{F2F0F0}777} & 2                    & 1138                 \\ \hline
\multicolumn{1}{l}{}                                 & \multicolumn{1}{l}{} & \multicolumn{1}{l}{}                             & \multicolumn{1}{l}{} & \multicolumn{1}{l}{}                             & \multicolumn{1}{l}{} & \multicolumn{1}{l}{}                             & \multicolumn{1}{l}{} & \multicolumn{1}{l}{} \\ \hline
\multicolumn{1}{c|}{}                                & \multicolumn{8}{c}{\textbf{Game B}}                                                                                                                                                                                                                                       \\ \cline{2-9} 
\multicolumn{1}{c|}{}                                & \multicolumn{2}{c|}{\textbf{TITAN}}                                     & \multicolumn{2}{c|}{\textbf{ReAct}}                                     & \multicolumn{2}{c|}{\textbf{Wuji}}                                      & \multicolumn{2}{c}{\textbf{Human}}          \\ \cline{2-9} 
\multicolumn{1}{c|}{\multirow{-3}{*}{\textbf{Task}}} & \#Bug                & \multicolumn{1}{c|}{Time (m)}                    & \#Bug                & \multicolumn{1}{c|}{Time (m)}                    & \#Bug                & \multicolumn{1}{c|}{Time (m)}                    & \#Bug                 & Time (m)             \\ \cline{2-9} 
\multicolumn{1}{c|}{1}                               & 0                    & \multicolumn{1}{c|}{2}                           & 0                    & \multicolumn{1}{c|}{5}                           & 0                    & \multicolumn{1}{c|}{4}                           & 0                    & 4                    \\
\multicolumn{1}{c|}{2}                               & 0                    & \multicolumn{1}{c|}{4}                           & 0                    & \multicolumn{1}{c|}{6}                           & 0                    & \multicolumn{1}{c|}{9}                           & 0                    & 6                    \\
\multicolumn{1}{c|}{3}                               & 1                    & \multicolumn{1}{c|}{4}                           & 1                    & \multicolumn{1}{c|}{4}                           & 1                    & \multicolumn{1}{c|}{12}                          & 0                    & 6                    \\
\multicolumn{1}{c|}{4}                               & 0                    & \multicolumn{1}{c|}{3}                           & 0                    & \multicolumn{1}{c|}{7}                           & 0                    & \multicolumn{1}{c|}{7}                           & 0                    & 9                    \\
\multicolumn{1}{c|}{5}                               & 1                    & \multicolumn{1}{c|}{7}                           & 0                    & \multicolumn{1}{c|}{13}                          & 1                    & \multicolumn{1}{c|}{9}                           & 1                    & 10                   \\
\multicolumn{1}{c|}{6}                               & 1                    & \multicolumn{1}{c|}{4}                           & 0                    & \multicolumn{1}{c|}{9}                           & 0                    & \multicolumn{1}{c|}{6}                           & 0                    & 10                   \\
\multicolumn{1}{c|}{7}                               & 1                    & \multicolumn{1}{c|}{9}                           & 0                    & \multicolumn{1}{c|}{14}                          & 1                    & \multicolumn{1}{c|}{9}                           & 1                    & 12                   \\
\multicolumn{1}{c|}{8}                               & 1                    & \multicolumn{1}{c|}{8}                           & 1                    & \multicolumn{1}{c|}{16}                          & 1                    & \multicolumn{1}{c|}{12}                          & 1                    & 11                   \\
\multicolumn{1}{c|}{9}                               & 1                    & \multicolumn{1}{c|}{12}                          & 0                    & \multicolumn{1}{c|}{18}                          & 0                    & \multicolumn{1}{c|}{19}                          & 0                    & 15                   \\
\multicolumn{1}{c|}{10}                              & 0                    & \multicolumn{1}{c|}{10}                          & 0                    & \multicolumn{1}{c|}{22}                          & 0                    & \multicolumn{1}{c|}{15}                          & 0                    & 17                   \\ \hline
\rowcolor[HTML]{F2F0F0} 
\multicolumn{1}{c|}{\cellcolor[HTML]{F2F0F0}Total}   & 6                    & \multicolumn{1}{c|}{\cellcolor[HTML]{F2F0F0}63}  & 2                    & \multicolumn{1}{c|}{\cellcolor[HTML]{F2F0F0}114} & 4                    & \multicolumn{1}{c|}{\cellcolor[HTML]{F2F0F0}102} & 3                    & 100                  \\ \hline
\end{tabular}
}
\end{table}

\begin{table}[t]
\centering
\caption{Distribution of identified bugs across different categories.}
\label{tab:bug-dist}
\resizebox{.7\textwidth}{!}{
\begin{tabular}{c|cccc|cccc}
\hline
                                    & \multicolumn{4}{c|}{\textbf{Game A}}                             & \multicolumn{4}{c}{\textbf{Game B}}                              \\ \cline{2-9} 
\multirow{-2}{*}{\textbf{Bug Type}} & \textbf{TITAN} & \textbf{ReAct} & \textbf{Wuji} & \textbf{Human} & \textbf{TITAN} & \textbf{ReAct} & \textbf{Wuji} & \textbf{Human} \\ \hline
Crash                               & 5              & 4              & 5             & 2              & 6              & 5              & 2             & 1              \\
Hang                                & 2              & 1              & 0             & 0              & 0              & 0              & 0             & 0              \\
Logic                               & 2              & 0              & 0             & 0              & 0              & 0              & 0             & 0              \\ \hline
\rowcolor[HTML]{F2F0F0} 
Total                               & 9              & 5              & 5             & 2              & 6              & 5              & 2             & 1              \\ \hline
\end{tabular}
}
\end{table}

\subsection{RQ3: Ablation Study}

To evaluate the contribution of \titan's core components to task execution and bug detection in MMORPGs, we conducted an ablation study under different configurations. Specifically, we focus on the three optimizable core components: the Perception Abstraction Module, the Action Optimization Module, and the Reflective Reasoning Module. All variants were coupled with the Oracle Detector, as bug identification would otherwise not be possible. We designed the following configurations: \textbf{TITAN}: full system with all three components enabled; \textbf{TITAN-A}: without the Perception Abstraction Module; \textbf{TITAN-O}: without the Action Optimization Module; \textbf{TITAN-R}: without the Reflective Reasoning Module. To validate the effectiveness of each component, we executed these variants on Game A across five independent runs. The detailed results are presented in Table~\ref{tab:rq3}.


Overall, the results show that each \titan component is highly synergistic. Removing any one module led to a substantial drop in performance, while removing all three—effectively reducing \titan to a basic LLM agent—resulted in performance comparable to the ReAct baseline, with only about 83\% task success and 45\% bug detection. These findings support our initial hypotheses and the motivation of our design: \ding{182} Without state abstraction, the state space is too complex, \ding{183} without action regulation, the action space becomes overwhelming \ding{184} without reflection, long tasks cannot be handled reliably. By addressing all of these challenges together, \titan achieves its strong overall performance.

\begin{tcolorbox}[size=title,opacityfill=0.1]
\noindent
\textbf{Finding 3:} Each core component of \titan contributes critically to its task success and bug‐detection performance. In particular, removing any one of the parameters can lead to substantial drops by up to 24\% in task completion and 27\% bug detection. The full \titan system outperforms all ablated variants by a clear margin, confirming the synergy among these components, especially in complex MMORPG tasks with long horizons.  
\end{tcolorbox}

\begin{table}[t]
\centering
\caption{Results of the ablation study for components in \titan.}
\label{tab:rq3}
\resizebox{1\textwidth}{!}{
\begin{tabular}{c|cccc|cccc|cccc|cccc}
\hline
                                & \multicolumn{4}{c|}{\textbf{TITAN}}  & \multicolumn{4}{c|}{\textbf{TITAN-A}} & \multicolumn{4}{c|}{\textbf{TITAN-O}} & \multicolumn{4}{c}{\textbf{TITAN-R}} \\ \cline{2-17} 
\multirow{-2}{*}{\textbf{Task}} & SR (\%) & CV (\%) & \#Bug & Time (m) & SR (\%)  & CV (\%) & \#Bug & Time (m) & SR (\%)  & CV (\%) & \#Bug & Time (m) & SR (\%) & CV (\%) & \#Bug & Time (m) \\ \hline
1                               & 100     & 82.11   & 1     & 129      & 60       & 80.55   & 1     & 40       & 80       & 86.15   & 1     & 168      & 80      & 55.61   & 1     & 12       \\
2                               & 100     & 76.25   & 1     & 135      & 100      & 39.71   & 0     & 24       & 100      & 30.95   & 1     & 128      & 100     & 64.45   & 1     & 30       \\
3                               & 100     & 50      & 1     & 148      & 100      & 48.5    & 1     & 33       & 100      & 45.8    & 1     & 152      & 100     & 32.3    & 1     & 12       \\
4                               & 100     & 57.26   & 1     & 64       & 80       & 62.96   & 0     & 89       & 100      & 49.36   & 1     & 78       & 100     & 53.98   & 1     & 33       \\
5                               & 100     & 93.43   & 0     & 63       & 80       & 86.63   & 0     & 66       & 100      & 80.13   & 0     & 55       & 100     & 36.67   & 0     & 67       \\
6                               & 100     & 64.5    & 0     & 42       & 100      & 31.7    & 0     & 80       & 100      & 28.2    & 0     & 48       & 100     & 96.4    & 0     & 58       \\
7                               & 100     & 68.75   & 1     & 53       & 60       & 80.85   & 1     & 92       & 80       & 78.65   & 0     & 57       & 80      & 47.13   & 0     & 45       \\
8                               & 80      & 68.61   & 0     & 16       & 60       & 64.91   & 0     & 205      & 80       & 62.81   & 0     & 24       & 80      & 39.9    & 0     & 113      \\
9                               & 100     & 84.55   & 2     & 36       & 80       & 81.35   & 2     & 162      & 100      & 78.35   & 2     & 33       & 100     & 63.58   & 2     & 147      \\
10                              & 80      & 76.31   & 2     & 18       & 0        & 74.41   & 1     & 224      & 40       & 71.71   & 2     & 20       & 60      & 58.53   & 1     & 133      \\ \hline
\rowcolor[HTML]{F2F0F0} 
Avg./Total                      & 96      & 72.17   & 9     & 704      & 72       & 65.16   & 6     & 1015     & 88       & 61.21   & 8     & 763      & 90      & 54.86   & 7     & 650      \\ \hline
\end{tabular}
}
\end{table}

\section{Discussion}

\subsection{Newly Detected Bugs}

During our evaluation, \titan uncovers four previously unknown bugs in the selected MMORPG task scenarios. We analyze three representative categories, demonstrating how \titan's architectural design enables the discovery of bugs that elude conventional testing methods. 

\begin{itemize}
    \item \textbf{Model Logic Bug}: \titan identified a severe logic bug where the character could clip through world geometry (e.g., walls and floors) into an unrendered void. This was traced to a missing asset that prevented collision data from loading. The LLM-guided agent's willingness to deviate from prescribed task execution paths and attempt unconventional interactions is crucial in surfacing this bug. In contrast, Wuji, ReAct, and human testers, who follow expected task execution routes and visible areas, never encounter this invisible-boundary issue, thus failing to expose this critical flaw in world integrity.
    \item \textbf{Hang Interaction Bug}: \titan also discovers an interaction bug that causes the game to hang under a specific condition. A particular item can be used or delivered via two interfaces: a direct on-screen UI button or the inventory menu. The DRL-based Wuji agent consistently uses the inventory menu method and therefore never exposes the problem. \titan, however, leverages its holistic perception module to identify all interactable elements on screen.  Its LLM-based action generation is not bound to a single optimal path and thus naturally explores alternative, valid interaction sequences.  This diverse action generation led it to use the on-screen button, which triggered an unhandled UI state transition and froze the game.  This class of bug, stemming from redundant but faulty UI paths, is systematically missed by agents with rigid interaction strategies.   
    \item \textbf{Step Counting Bug}: \titan detected a subtle quest logic error where a task requiring six enemy kills was marked complete after only five. This off-by-one error was missed by all baselines. Human testers and deterministic agents strictly follow instructions, killing all six targets and observing correct behavior. \titan's dynamic and reactive planning capabilities lead to this discovery. For instance, the agent might pause its combat task to engage a newly spawned resource node before returning to the quest objective. This non-linear, interrupt-driven behavior created an execution trace that exposed the faulty completion condition in the quest's state machine. 
\end{itemize}

\subsection{Implications}

The development of \titan and our comprehensive evaluation offer several important implications for both the research community and the game industry, particularly in addressing the critical challenges of cost, efficiency, and adaptability in modern game testing.
\begin{itemize}
    \item \textbf{Prerequisite of Task Completion for Practical Adoption.} The substantial cost of game testing, often consuming nearly half of a development budget, forces many studios to severely limit regression testing, sometimes to only a single full run before a release. In this context, the primary value of an automated testing agent is not merely its ability to explore or its bug-finding potential, but first and foremost its ability to reliably complete core game tasks. An agent that frequently gets stuck or fails to finish a game task cannot be integrated into a high-stakes production pipeline, no matter how clever its exploration strategy might be. \titan's high task success rate, which approaches human-level reliability, demonstrates that LLM-driven agents can achieve the necessary level of robustness to serve as a foundation for automated regression testing. This capability can significantly reduce the manual burden on human testers, allowing them to focus on more creative and complex test scenarios that truly require human intuition.

    \item \textbf{The Need for Adaptive MMORPG Testing Solutions.} The game industry is characterized by frequent and rapid updates, often with multiple daily builds. This pace renders traditional automation approaches economically unsustainable. Rule-based scripts require constant, expensive manual maintenance to adapt to even minor game changes. DRL-based agents, while more adaptive, require costly per-task retraining and reward engineering, making them impractical for newly introduced content or frequently evolving mechanics. \titan presents a compelling alternative: a training-free framework that leverages the zero-shot reasoning and in-context learning capabilities of LLMs. By using game-agnostic abstraction templates and RAG-retrieved knowledge, \titan can be rapidly configured for new games or new content within minutes, as demonstrated by its application to two distinct MMORPGs. This agility addresses a critical industry need for testing solutions that can keep pace with modern development cycles without incurring prohibitive costs.

    \item \textbf{Towards Comprehensive and Holistic Test Automation.} Beyond completing tasks, \titan demonstrates that a single agent framework can simultaneously address multiple testing objectives: functional correctness (via task completion), thoroughness (via coverage-guided exploration), and bug detection (via advanced oracles). This holistic approach moves the field beyond tools that excel in only one dimension. The framework's ability to uncover novel bug types, such as logic errors and soft locks, that are often missed by other methods and human testers, highlights its potential to improve overall software quality and not just testing efficiency. The integration of LLM-based reflection and diagnosis further reduces the triage burden on developers by providing contextualized bug reports, shifting the role of automation from mere task execution to intelligent analysis. Similar concerns for systematic test optimization and robustness have also been emphasized in the DL testing literature~\cite{hu2024test,hu2025assessing}, suggesting that insights from broader AI testing can inform comprehensive game testing.

\end{itemize}





\subsection{Threats to Validity}

\textbf{Internal Validity.} The non-deterministic nature of LLMs introduces stochasticity into \titan's decision-making. Consequently, specific bugs or execution results cannot be guaranteed to generalize across all trials. To mitigate this, we conduct all experiments over five trials and incorporate coverage-guided exploration to stabilize overall performance. Due to cost and time constraints, we use only OpenAI GPT-4o as the foundation model. However, \titan is designed to support other foundation models as well. For prompt design and parameter settings, we follow empirical guidance and best practices from our collaborating company to ensure that the results are as strong and consistent as possible.

\textbf{External Validity.} The generalizability of our findings may be limited to MMORPGs and similar genres with rich narratives and game task-based structures. \titan's performance may diminish in genres that rely heavily on reflexes or minimalistic narratives, where LLM-based reasoning provides fewer advantages. Furthermore, the agent's effectiveness depends partly on the alignment between game mechanics and the LLM's pre-existing knowledge of conventional game logic. Our evaluation is restricted to two games due to time and budget constraints, although \titan has since been deployed in production environments of several commercial games of our collaborating game companies, suggesting its broader applicability.

\section{Conclusion}

In this paper, we present \titan, the first LLM-driven agent framework for automated testing of MMORPGs. \titan integrates state abstraction, action regulation, reflective reasoning, and diagnostic oracles to address the high costs and limited reasoning capabilities of prior methods. Our evaluation on two commercial games shows that \titan outperforms state-of-the-art baselines and human testers in both task completion and bug detection. It successfully identifies four previously unknown logic and performance bugs that prior approaches miss. Ablation studies further confirm the essential contribution of each component to \titan's overall effectiveness.

\titan offers a practical, training-free solution that adapts quickly to game changes, making it suitable for integration into real-world QA pipelines. More broadly, our work demonstrates the promise of structured LLM-based agents for complex game testing and opens new avenues for advancing intelligent, general-purpose testing systems.

\bibliographystyle{unsrtnat}
\bibliography{references}  







\end{document}